Coupled Loop-within-Loop Resonators at the near-field: Extended Bandwidth and Short Pulse Response upon Parametric Conversion

H. Grebel

Electronic Imaging Center and the department of Electrical and Computer Engineering, NJIT, Newark, NJ, 07102. grebel@njit.edu

**Abstract:** Parametric frequency conversion have seen a renewed interest, specifically in superconducting quantum circuits at GHz range. These nonlinear convertors are typically narrow band and one may wish to extend their frequency range. Here, two coupled resonators are investigated: a loop-within-loop structure, whose gap is coupled by short, a diode, or simply left open. While simulations and experimental results are limited to GHz range, the concept is general. On their own, the loops exhibit natural frequencies of ca, f≈1.5 GHz and f/2≈0.73 GHz, respectively; the coupling modifies the resonance frequencies and increases its bandwidth. In one example, signal is fed from one loop and is received by the other. In another example, the input signal is provided across a coupling diode and is received by either loops. It is shown that in the case of nonlinear coupling, the frequency bandwidth can be as large as 3 GHz. Such bandwidth is associated with a very narrow temporal response of less than 150 ps, suggesting coherent effects through high-harmonic excitations at the near-field.

## I. Introduction

Ultra-short pulses (USP) are well-known for optical applications, such as, laser ablation, information and communication systems, and range detection [1-2]. RF/microwave pulse applications, such as radar, ranging and electronic warfare are also known, but are susceptible to dispersion limitations and limited temporal coherence.

Parametric conversions are nonlinear effects. In optics, energy transfer to multiple harmonics, most notably, second and third harmonics, or half-frequencies are enabled by nonlinear crystals [3]. In the RF/microwave regime, nonlinear varactors, diodes, or memristors [4-7] were suggested for these parametric conversions. Quantum approach [8] to superconducting circuits makes heavy use of these conversions [9-14]. As currently constructed, the amplification and oscillations occurs at relatively narrow frequency bands. We investigate here the possibility of a broadband parametric conversion in the RF/microwaves and specifically, loop resonators - loop resonators present convenient, conformable and planar structures which are amenable to self-contained magnetic flux and near-field confinement.

In general, the resonance frequency of a medium-size, single and thin loop depends on the loop material, its thickness (say, a 35 micron copper film) and its substrate (say, epoxy based printed circuit board, PCB). For example, a thin loop with a diameter of 8 cm exhibits a resonance of ca 1 GHz with a bandwidth of ca 150 MHz. This center frequency changes as the loop width increases. Two relatively thick resonators, say a loop-within-loop, exhibit mutual capacitance and mutual inductance that in addition of a physical coupling element (e.g., a linear resistor, capacitor, or a nonlinear diode) affect the final resonance values.

Here, two coupled resonators, a loop-within-loop structure, whose gap is coupled by short, a diode, or simply left open, are investigated by simulations and experiments. The elements exhibit a broad band of frequencies as a sum and difference of the basic natural resonating frequencies. In the diode case, the structure exhibits bandwidth over 3 GHz, which is correlated with its Fourier transformed, pulse-width of less than 150 ps. Such a broad band effect suggests coherent higher-harmonics contributions. Simulations for the structures with a monocycle are provided in Section II. Experiments using VNA are provided in Section III: Discussions and Conclusions are provided in Section IV.

## II. Simulations

Simulations were conducted using a Comsol Multiphysics code. A physics-controlled normal mesh was used; finer meshes did not result in much a better accuracy; a built-in PML sphere in the time domains was situated at an outer radius of 20 cm and thickness of 3 cm (amounting to the lowest wavelength) with enough distance from the structure; polynomial coordinate stretching and scaling/curvature parameters of 1/1, were used.

The structure is illustrated in Fig. 1. Two concentric loops are physically coupled by either a diode, a short, or through mutual capacitance and inductance when the gap between them is left open. Copper loops of thickness 35 microns, are situated on printed circuit board (relative dielectric constant of $\varepsilon_r$=4). Their natural frequencies were f≈1.5 GHz (high-frequency, HF) and low frequency (LF) f/2≈0.73 GHz, respectively, as estimated by an analytical model using lump elements (not shown here). Size wise, the LF loop diameter was 8 cm with a width of 1 cm and the HF loop diameter was 4 cm with a width of 0.5 cm.

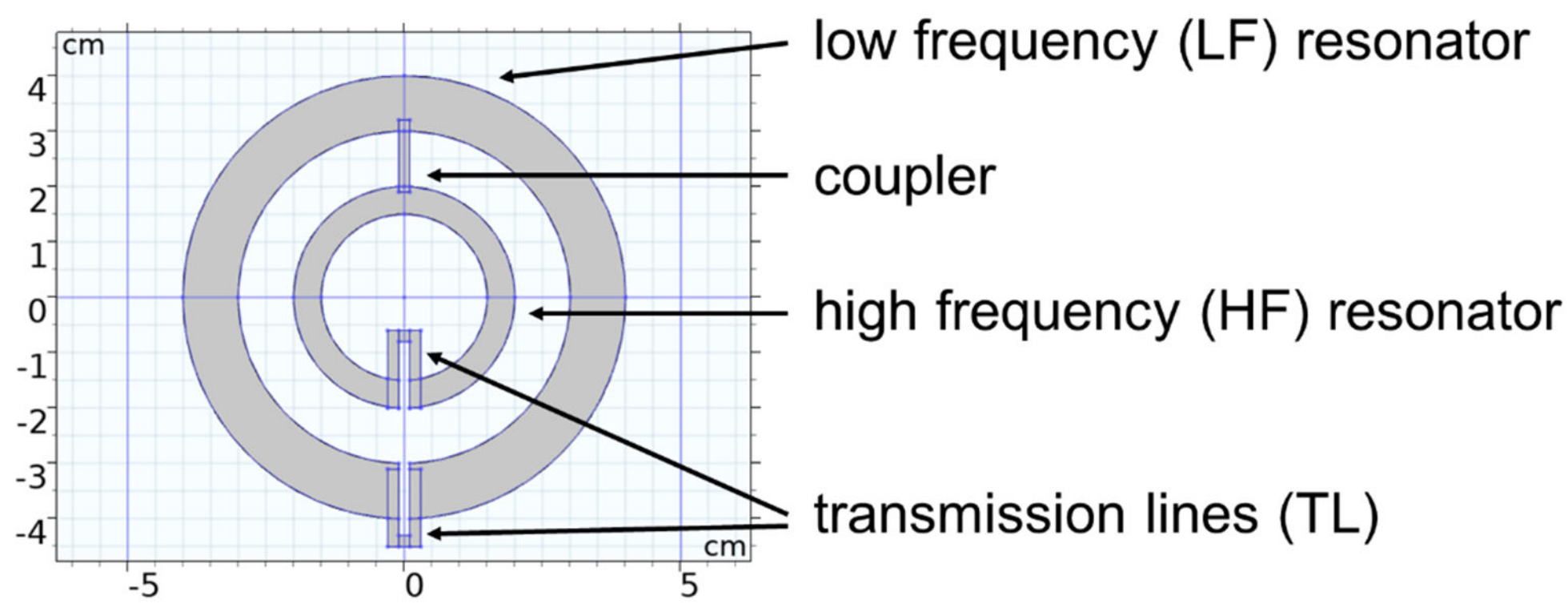

**Fig. 1.** The simulated structure: two concentric resonating loops are either coupled by a diode, a short, or left open. Input signal are provided by the HF transmission line (TL) and received by the TL of LF loop, or vice versa.

The open and short cases require only one module of Comsol's code (RF). For the Schottky diode simulations, one adds a circuit module with the following parameters:

Reverse breakdown, $B_v=\infty$; zero-biase junction capacitance, $C_{j0}=10^{-12}$ F; forward bias capacitance, $F_c$=0.5; current breakdown voltage, $I_{bv}=10^{-9}$ A; corner for high-current roll-off, $I_{ke}=\infty$; saturation current, $I_s=10^{-12}$ A; reverse saturation current, $I_{se}$=0 A; grading coefficient, M=0.5; ideality factor, N=1 (note: making N=1.05 as typical for Schottky diodes does not made much difference); breakdown ideality factor. $N_{bv}$=1; reverse ideality factor, $N_r$=2; series resistance, $R_s$=10 Ohms; Temperature, T=293 $^o$K; junction potential $V_j$=0.4. The activation of the diode is made through the voltage difference between the HF and LF loops via external I vs U option.

Comsol is not performing well for continuous, non-linear elements in the frequency domain. In order to circumvent this, we turned to simulations in the time domain with an input of a monocycle that has both positive and negative lobes and provides for a range of frequencies. A Fast Fourier Transform (FFT) was then performed to better compare it to the experimental results. Experimental results were obtained with a cosine signal in the frequency domain, using Vector Network Analyzer (VNA) by Copper Mountains Labs. Then, the results were transformed by IFFT to the time domain.

The simulated temporal signal - a Gaussian monocycle – had the form,

$$2 \cdot \left(\frac{A_S}{\tau}\right) \cdot \left(\sqrt{e}\right) \cdot (t - Tc) \cdot e^{-2[(t-Tc)/\tau]^2}). \quad (1)$$

Here: t is the time; $A_s$ is the signal amplitude, here, $A_s$=1; $\tau$ is the cycle width, in our case, $\tau$=0.5 ns; Tc is the delay, in our case, Tc=0.5 ns. The simulation results for the open, short and diode cases are shown below.

Terminology is as follows: HF/LF means that the monocycle signal was feeding the HF loop through the transmission line (TL) and received by the TL of the LF loop; LF/HF means the reverse.

<u>Open Case: the input signal</u>

This case is presented in Fig. 2. As stated before, the gap between the loops remained open. Couplings between loops is made through mutual capacitance and mutual inductance in the near-field. While the high-frequency (HF) input may invoke higher-harmonics in the low frequency loop (LF) the reverse may only partially true: low frequencies in the HF loop may not entirely cut-off and may propagate for short distances due to loop width and near-field effects.

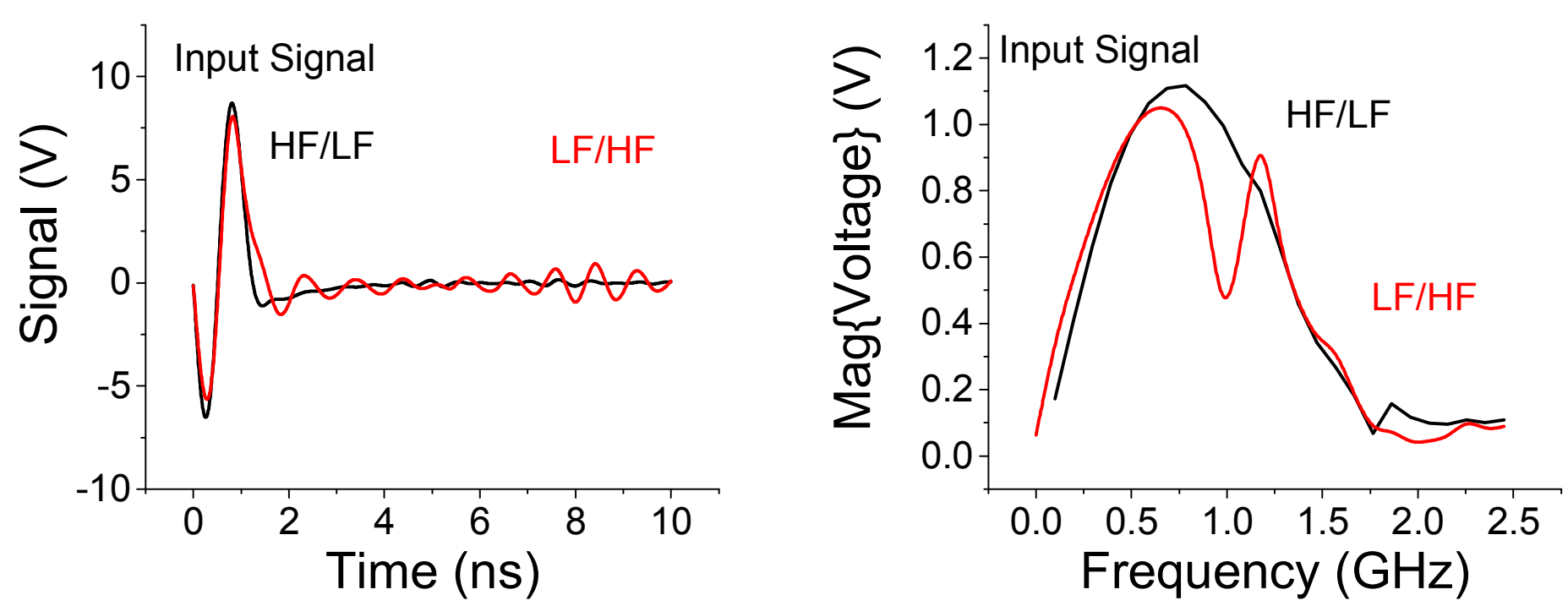

(a) (b)

**Fig. 2.** Open case. HF and LF input signals in the time (a) and frequency (b) domains. The HF spectrum exhibits a single peak while the LF exhibits two peaks near the high and low resonances. The deviation from the natural frequencies is attributed to the mutual capacitive and inductive couplings.

<u>Open Case: The received signal</u>

The received signals in the open case are presented in Fig. 3. The HF and LF spectra clearly exhibit multiple peaks in the region of 0.75-1.75 GHz and a very low-frequency peak at 0.25 GHz.

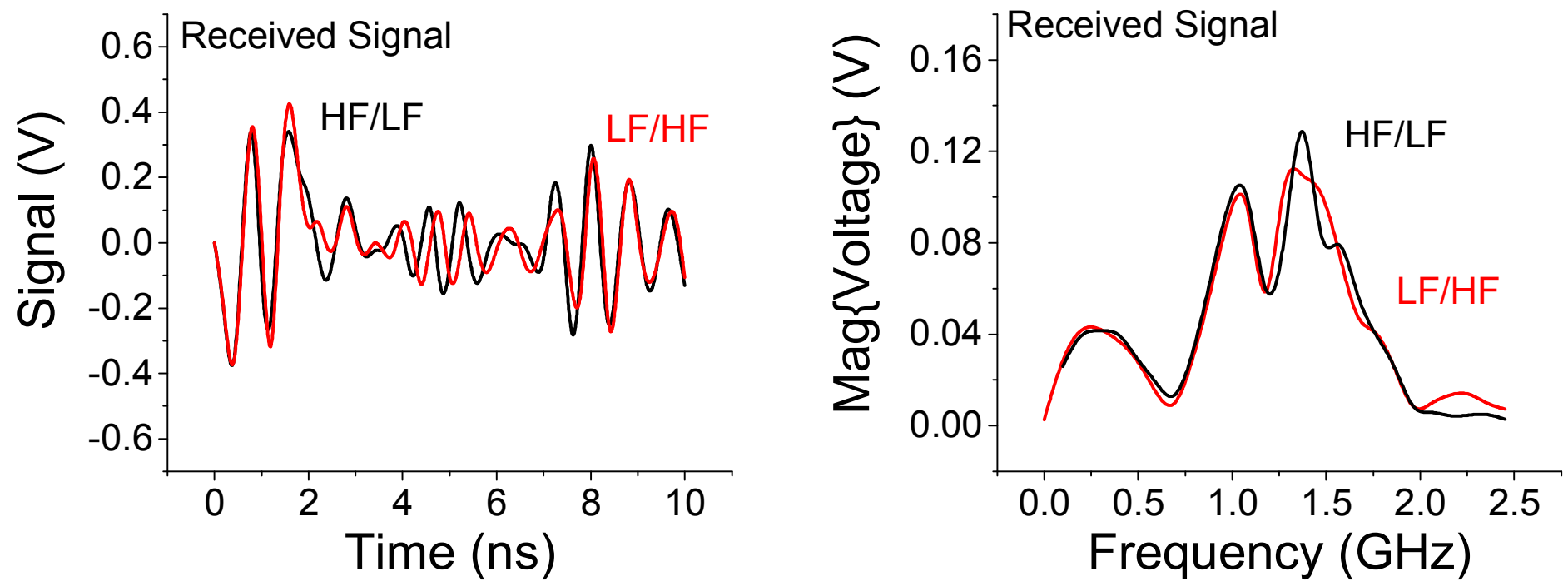

(a) (b)

**Fig. 3.** Open Case. HF and LF received signals in the time (a) and frequency (b) domains. Note the build-up of the resonance over time as shown by the delayed response. The HF and LF spectra exhibit multiple peaks in the region of 0.75-1.75 GHz and at the very low-frequency of 0.25 GHz.

<u>Short Case: the input signal</u>

The input signal in the Short case is presented in Fig 4. It is rather similar to the input for the Open case.

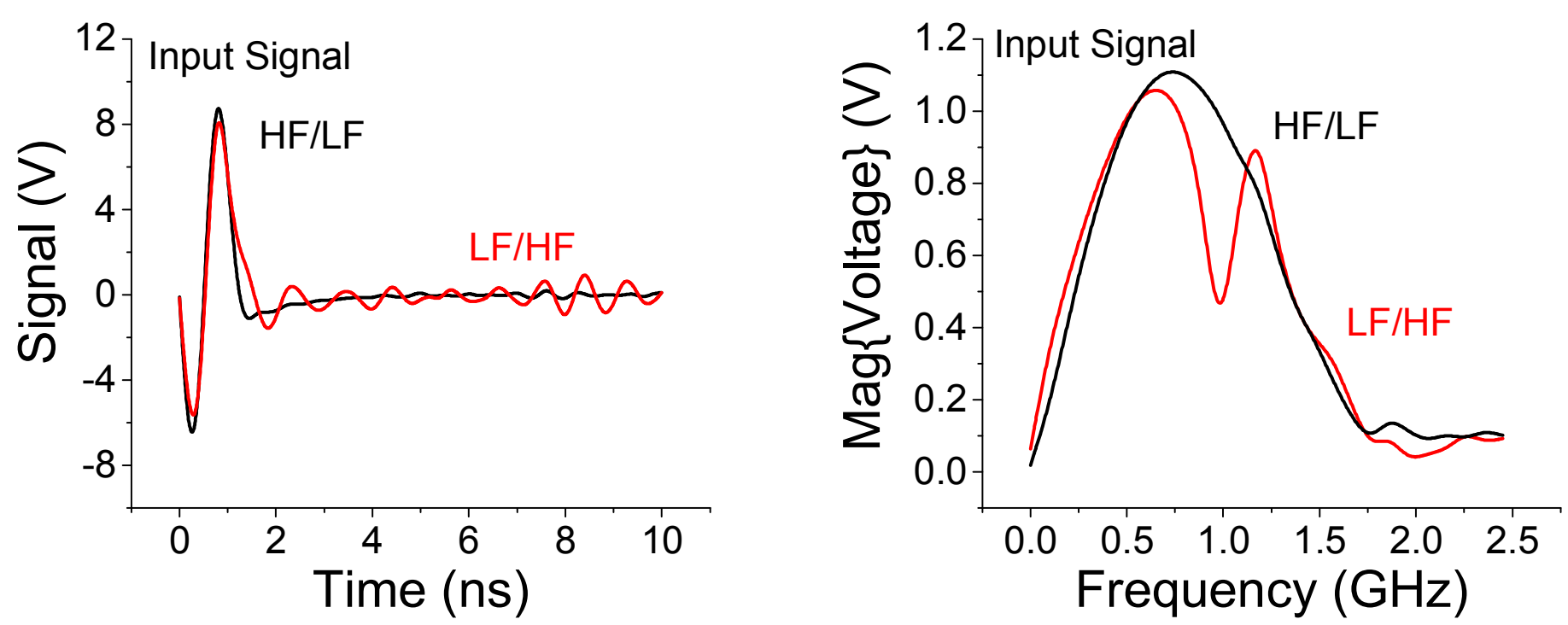

(a) (b)

**Fig. 4.** Short case. HF and LF input signals in the time (a) and frequency domains (b). The HF spectrum exhibits mostly single peak while the LF exhibits two peaks at the 0.25-1.25 GHz region.

<u>Short Case: the received signal</u>

The received signals for the short case are presented in Fig. 5. The HF and LF spectra clearly exhibit multiple peaks in the region of 0.75-1.75 GHz and a very low-frequency peak at 0.25 GHz. The peak in the 1.25-1.75 GHz has been split to two.

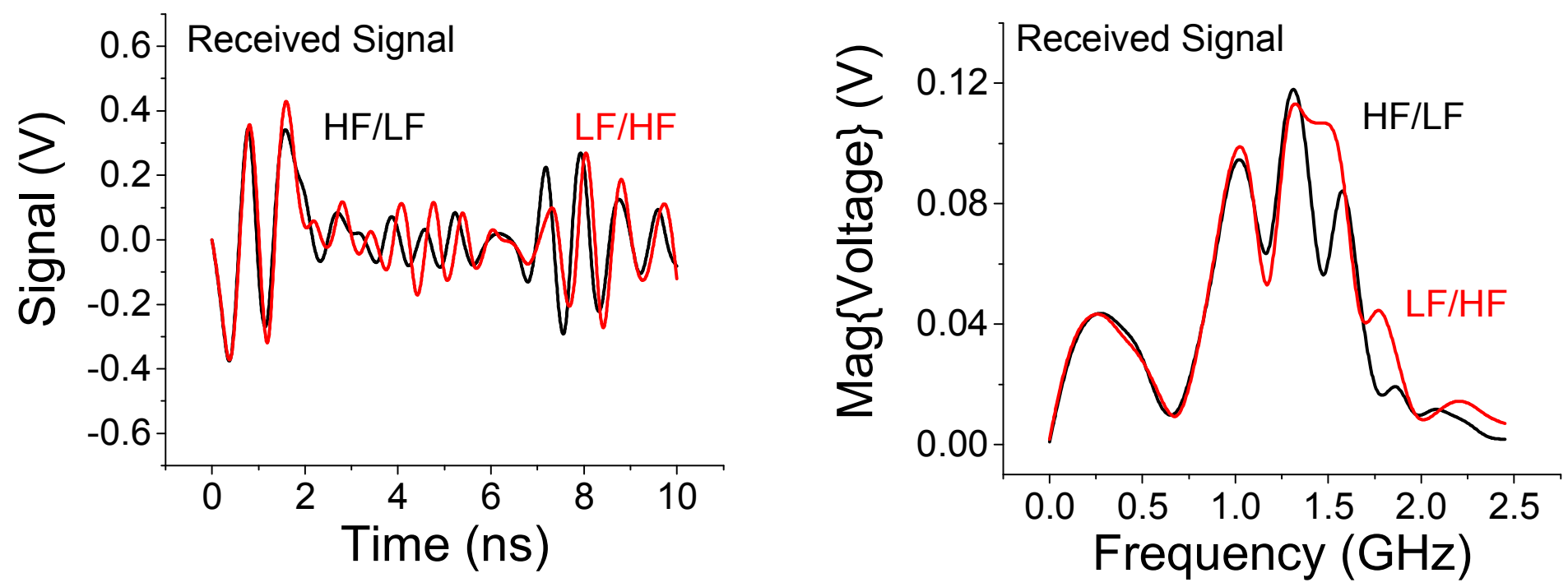

**Fig. 5.** Short case. HF and LF received signals in the time and frequency domains. The HF and LF spectra exhibit triple peak. The two peaks at 1.25 GHz and 1.6 GHz peaks are merged when the input is made through the TL of the LF loop.

Diode Case: the input signal

Here, coupling is provided by a diode. The diode polarity is from the HF loop to the LP loop. The corresponding phase linearity for when both loops are coupled by a diode whose polarity is from the High-frequency loop (HF) to the low frequency loop (LF) is shown in Fig. 6c for both the HF input and the receiving, LF output. Here, the linearity of the phase has shifted from an input range of 0.2-1.2 GHz to 0.75-1.75 GHz range for the output.

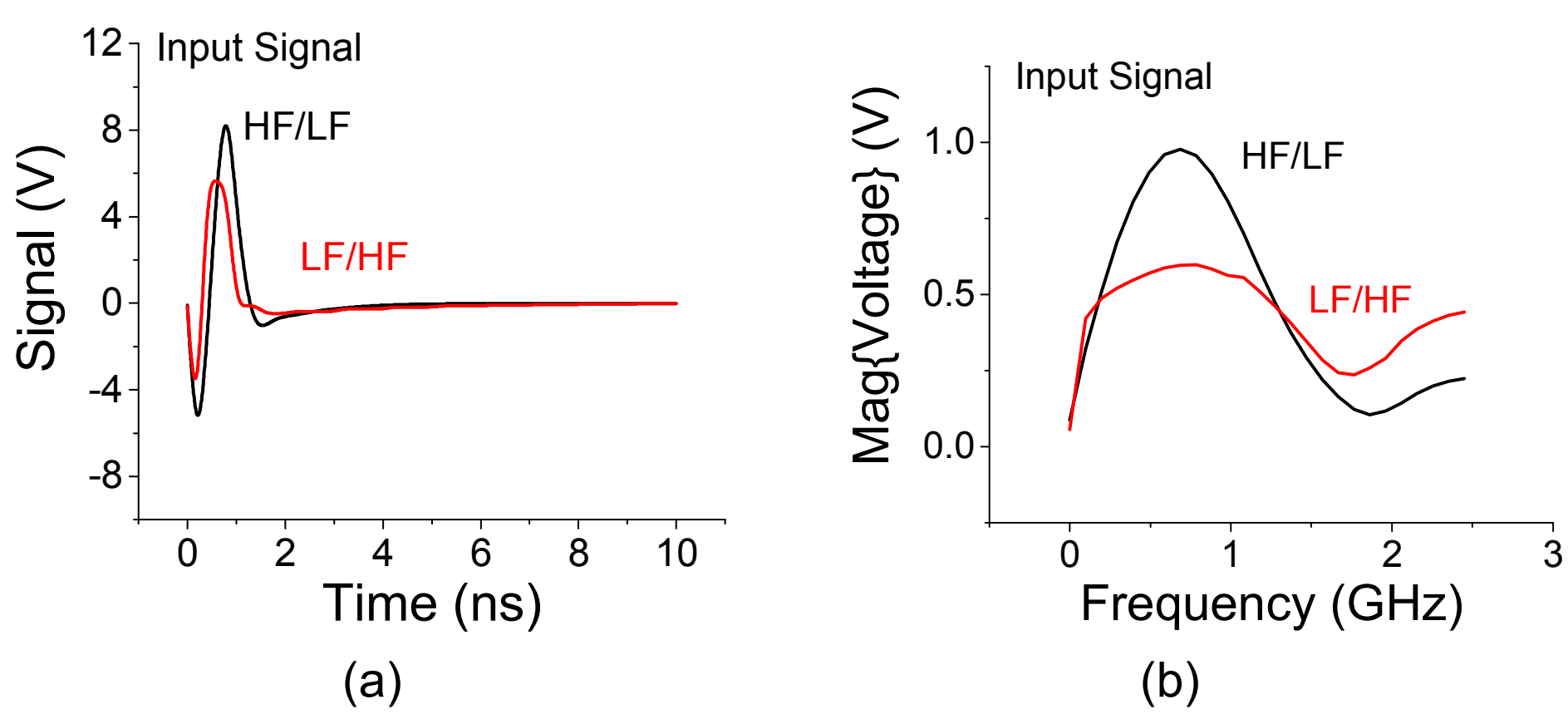

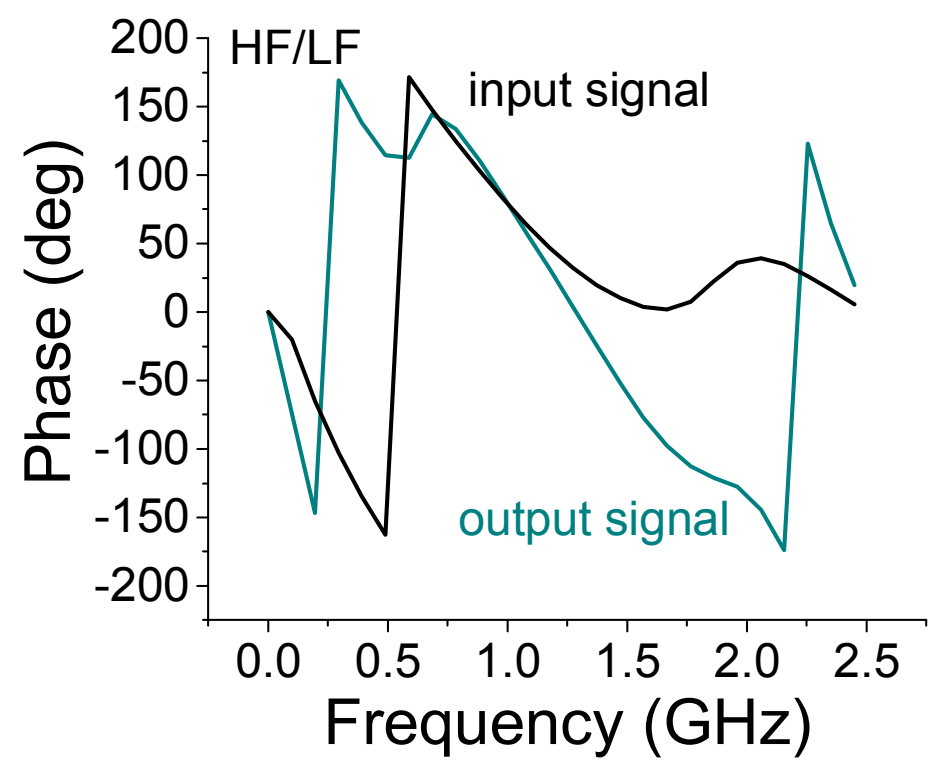

(c)

**Fig. 6.** Diode case. (a,b) HF input signals in the time and frequency domains for a diode coupler. The HF and LF spectra exhibited a single peak. (c) The corresponding phase linearity for the HF/LF case; the linearity of the phase has shifted from an input in 0.2-1.2 GHz region to 0.75-1.75 GHz for the output.

<u>Diode Case: the received signal</u>

The received signals are presented in Fig. 7. The output phase in the in the receiver (LF loop) is shown in Fig. 6c corroborating the linearity of the received signal. The high-frequency peak at 1.25 GHz has been extended, while the low frequency at 0.25 GHz has been broaden and its amplitude increased.

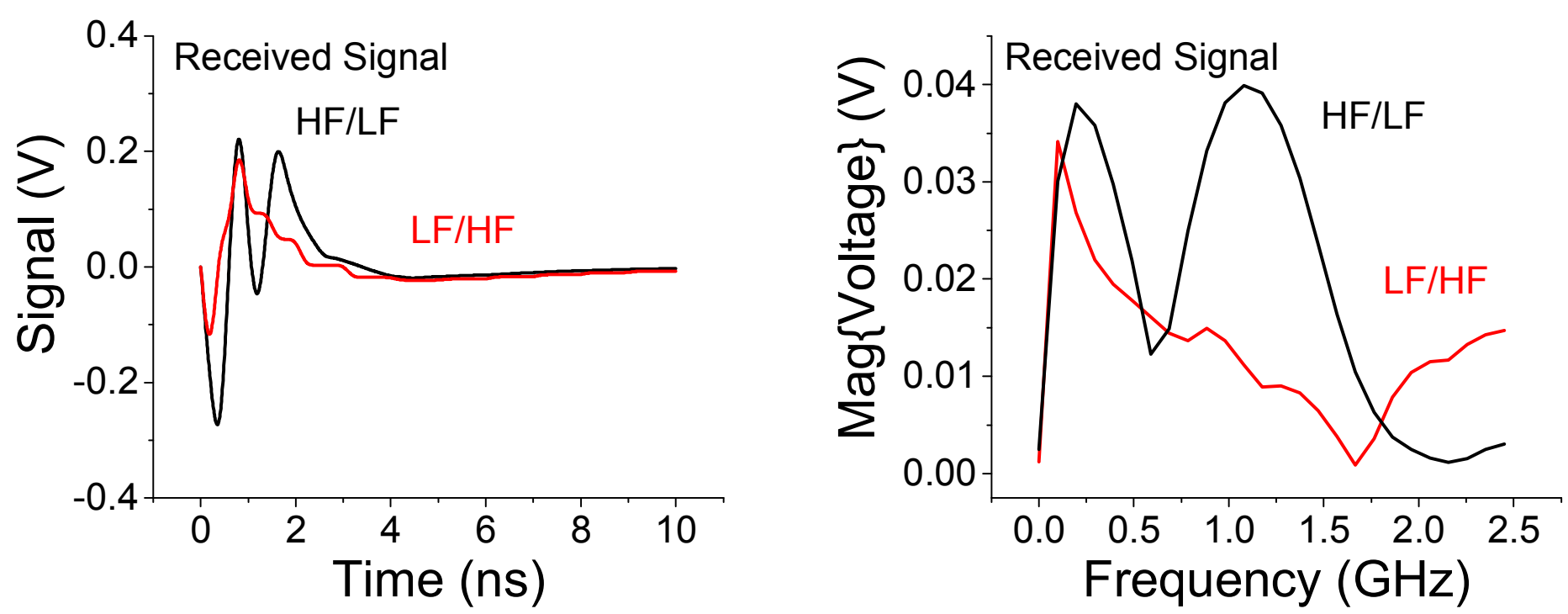

(a) (b)

**Fig. 7.** Diode case. HF and LF received signals in the time (a) and frequency (b) domains. The HF and LF spectra clearly exhibit double peak at high and low resonances. The deviation from the natural frequencies is attributed to the mutual capacitive and inductive coupling. The high-frequency peak at 1.25 GHz has been extended, while the low frequency at 0.25 GHz has been broaden and its amplitude increased.

Experiments were conducted with a floating metal frame around the structure. The frame might have created losses in the nearby LF loop. This, turned out, was not significant. The frame has not been affecting the results, substantially (Fig. 8). It is interesting to note that the diode itself exhibits a bit larger bandwidth (Full Width at Half Max, FWHM)) than the receiving port, 0.4-1.55 GHz while the linear phase is extended from 0.3-1.85 GHz.

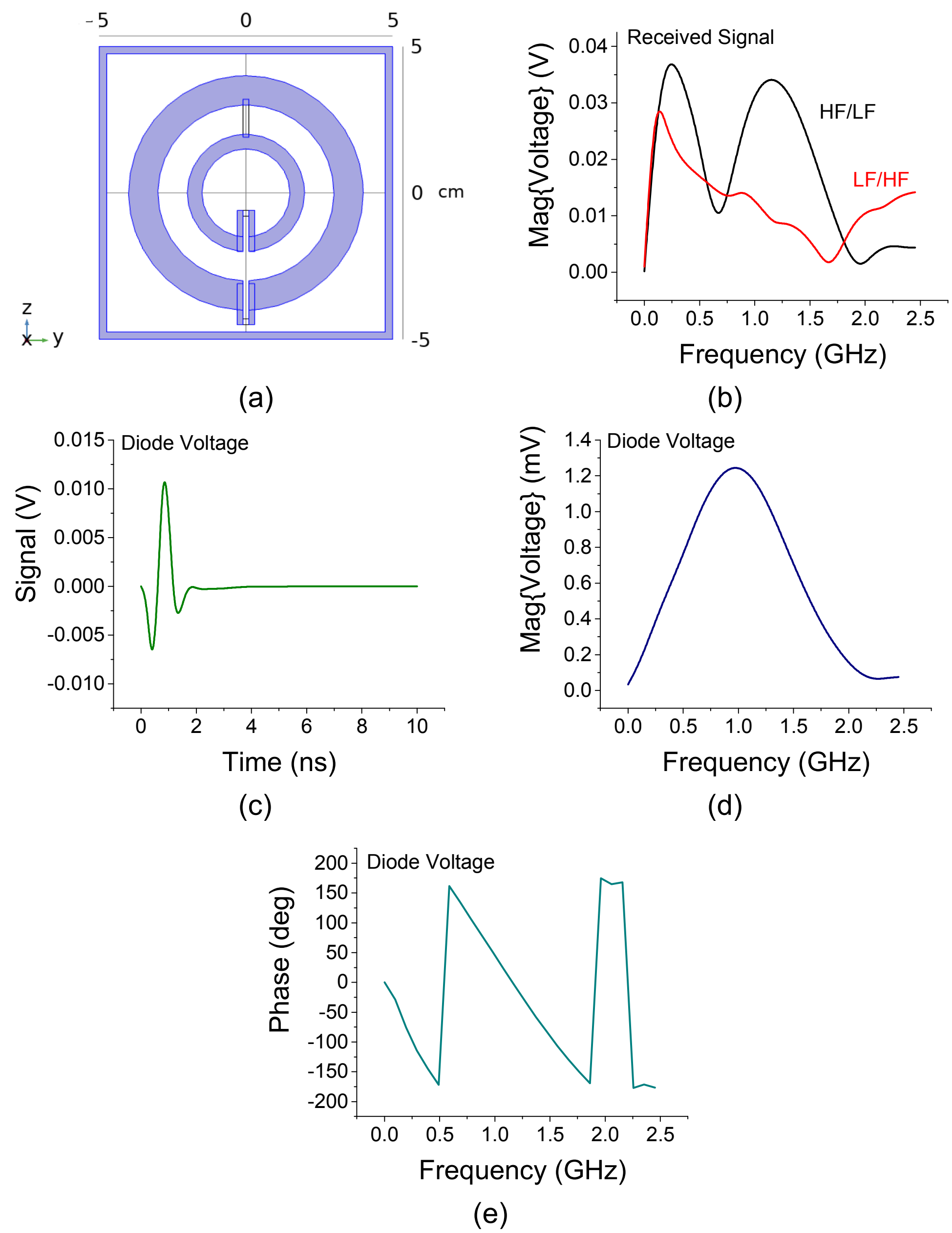

**Fig. 8.** Metal frame case. (a) A structure with metal frame. (b) HF/LF and LF/HF for a diode coupler. (c) Input voltage across the diode. (d) Diode bandwidth response. (e) The voltage phase.

Finally, one might ask what happens if we feed the signal across the diode itself (thereby, directly changing the coupling between loops). In this case one ought to examine the effect of either outputs: HF and LF. The direction of the diode should matter as exhibited in Fig. 9. The monocycle is fed through a TL in parallel to the diode and there is a resistor in series with the diode while both are parallel to the diode's TL.

When the diode is polarized from HF to LF, there is no change in the spectra upon changing the resistance (which is in series with the diode). The HF loop exhibits three peaks: at 0.75, 1.3 and 2.1 GHz. The LF loop exhibits two peaks in that region: 0.75 and 1.8 GHz. However, when the polarity of the diode is reversed (LF to HF), a shift in the loop's spectra is detected: for the HF loop, the 0.75 and its second harmonics 1.3 GHz peaks get closer; for the LF loop, the 0.75 and the third harmonics at 1.8 GHz become more separated.

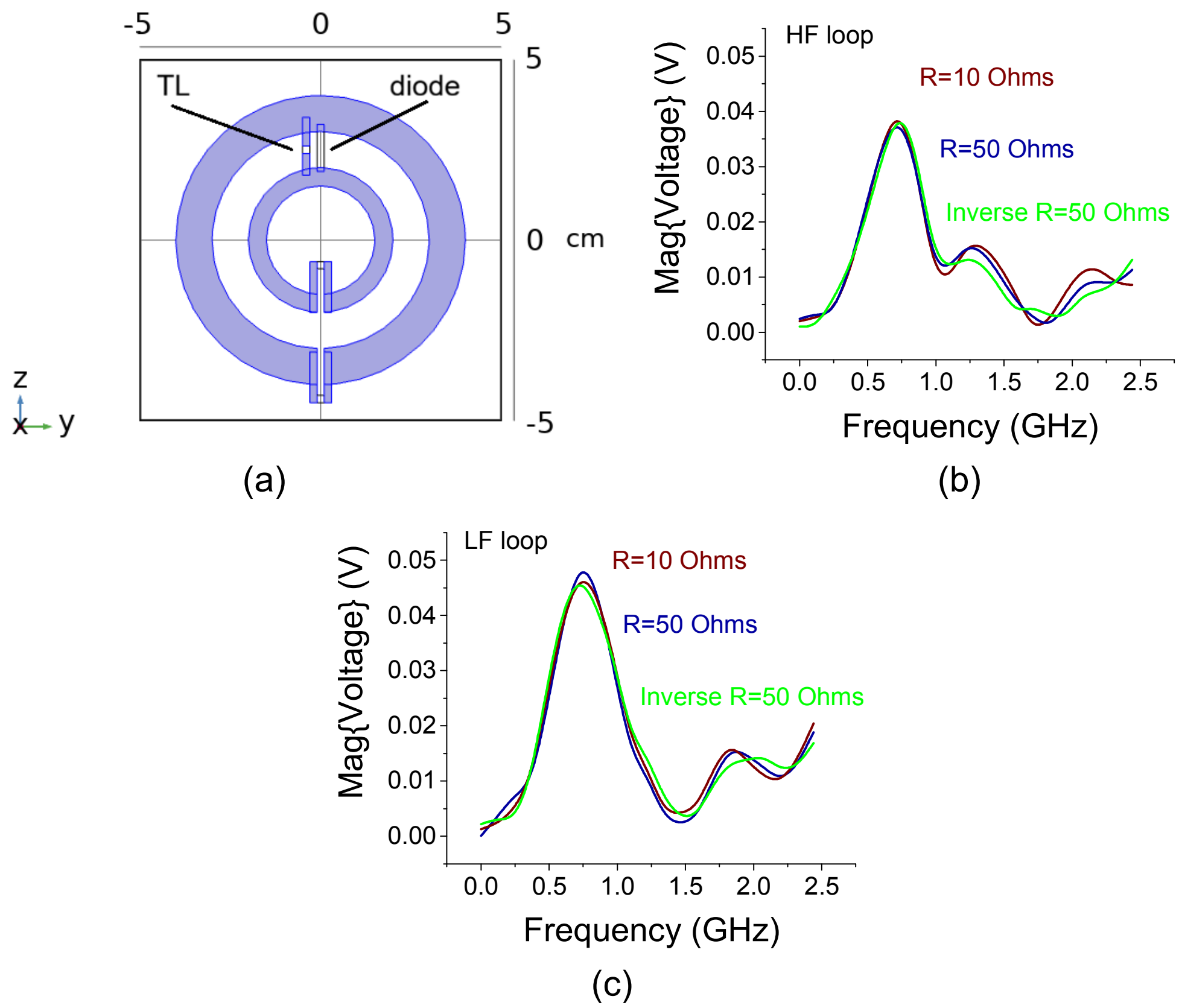

**Fig. 9.** Input across the diode. (a) Feeding the HF and LF loops by a monocycle directly through the diode. The diode is polarized from HF to LF and there is no change in the spectra when diode resistance is increased from 10 to 50 Ohms. The HF loop exhibits three peaks: at 0.75, 1.3 and 2.1 GHz. The LF loop exhibits two peaks in that region: 0.75 and 1.8 GHz. However, when the polarity of the diode is reversed (LF to HF) a shift in the loop's spectra is detected: for the HF loop, the 0.75 and its second harmonics 1.3 GHz peaks get closer; for the LF loop, the 0.75 and the third harmonics at 1.8 GHz become more separated

## III. Experiments

The structure was made by lift-off of dry etching photoresist on copper coated, standard printed circuit board. The soldering points are for SMA connectors at the board back. A Schottky diode (Mauser) that is rated at bandwidth of 10 GHz was attached to the loops. The Open case was measured without the diode and the Short case was measured with a copper tape that replaced the diode. A 1 MHz to 8 GHz VNA, calibrated with an accompanied standard cross, was used (Copper-Mountain Technologies). The system has a time domain capabilities via IFFT module, thus one is able to assess the coherence of the results.

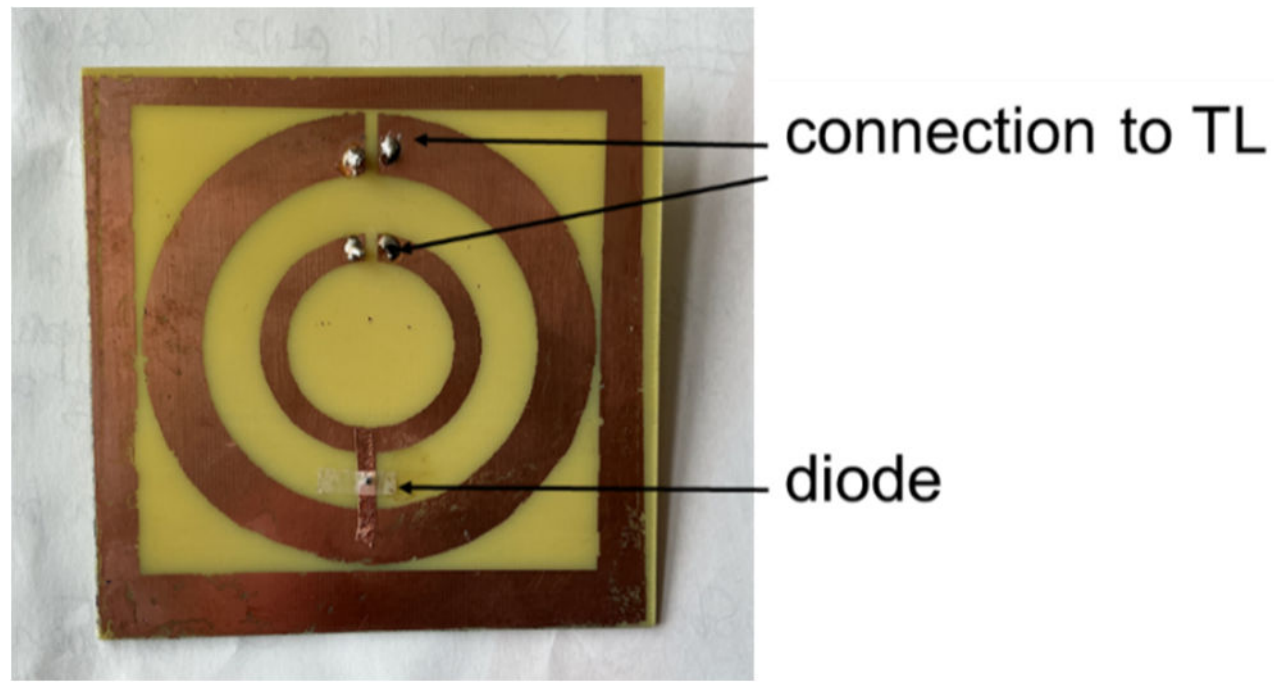

**Fig 10.** The structure with a diode, placed between the loops. Feed to each loop was made via SMA cables from the back. Short coupling is made by replacing the diode with a copper tape. The diode's polarity was ascertain with a Multimeter (MM).

The experimental curves are presented in Figs. 11, 12. Fig. 11a presents the entire (smoothed) curves for the HF/LF cases; the inset accentuates the 1-3 GHz region. Fig. 11b presents the equivalent responses in the time domain - the VNA is capable of IFFT and the point here is to show that the extended frequency bandwidth is associated with a corresponding decrease in pulse response. The inset accentuates the response in the 3.25-3.8 ns region to exhibit a temporal response of less than 150 ps. The horizontal line marks the -3 dB limit: the lines covers only one peak in the case of the Diode and two peaks the case of the Open and Short cases. In Fig. 11c, reverse launching, LF/HF is shown. In this case the broadband effect of the diode has disappeared. Fig. 11c shows that the launching the signal through the LF port and received by the HF port results in a much smaller frequency bandwidth.

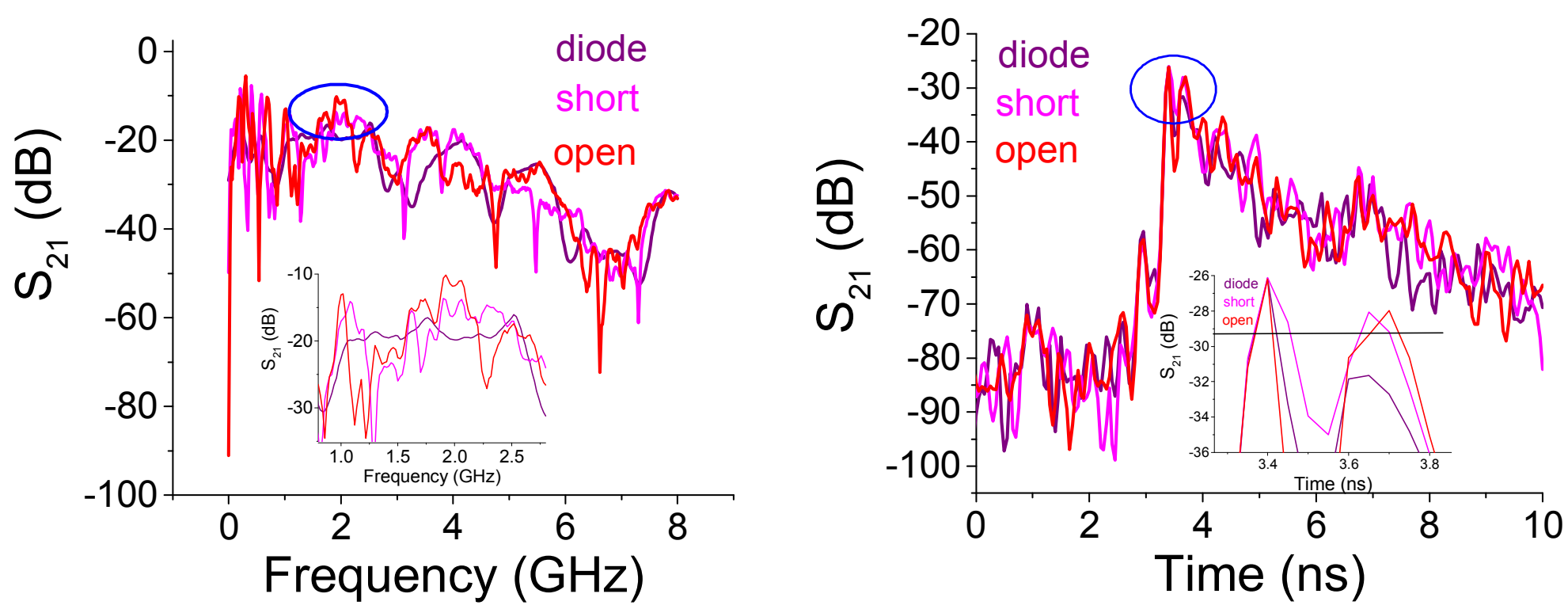

(a) (b)

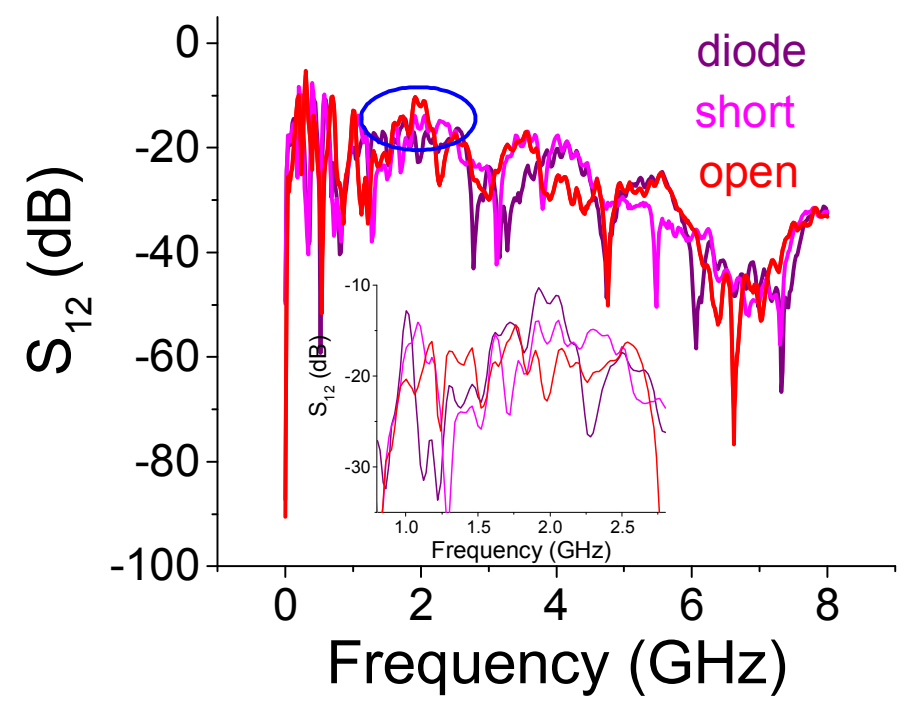

(c)

**Fig. 11.** Experiments. (a) Transmitted signal $S_{21}$ when signal is launched through the HF's TL and received by the LF's TL. It exhibits a bandwidth of ca 1.5 GHz (b) FFT of the frequency exhibits a coherent temporal response of less than 150 ps. The horizontal line points to -3 dB value: for the open and short coupling it covers both sub-peaks; for the diode it covers only one peak. (c) The signal is launched through the LF port and received by the HF TL. The solid lines point to accentuated region in the inset and reveals little bandwidth increase.

Experiments in the frequency domain when a signal is launched across the diode are shown in Fig. 12. The figures show two polarity cases: when the diode is forward bias from LF to HF (Fig. 12a.): here the frequency bandwidth is quite small. The case where the diode polarity is from the HF loop toward the LF loop is shown in Figs. 12b and 12c in the frequency and times domains, respectively. Similarly to the simulations, we note a small peak at 0.25 GHz. When the diode polarity is from the HF loop to the LF loop and the signal is received by the HF loop, there exist two relatively broad peaks around 2 and 5 GHz with broadened bandwidths. Even a larger bandwidth is received for the LP loop output (which is in reverse order to the diode polarity) - this 3-6 GHz band is associated with a quite narrow temporal response of less than 150 ps.

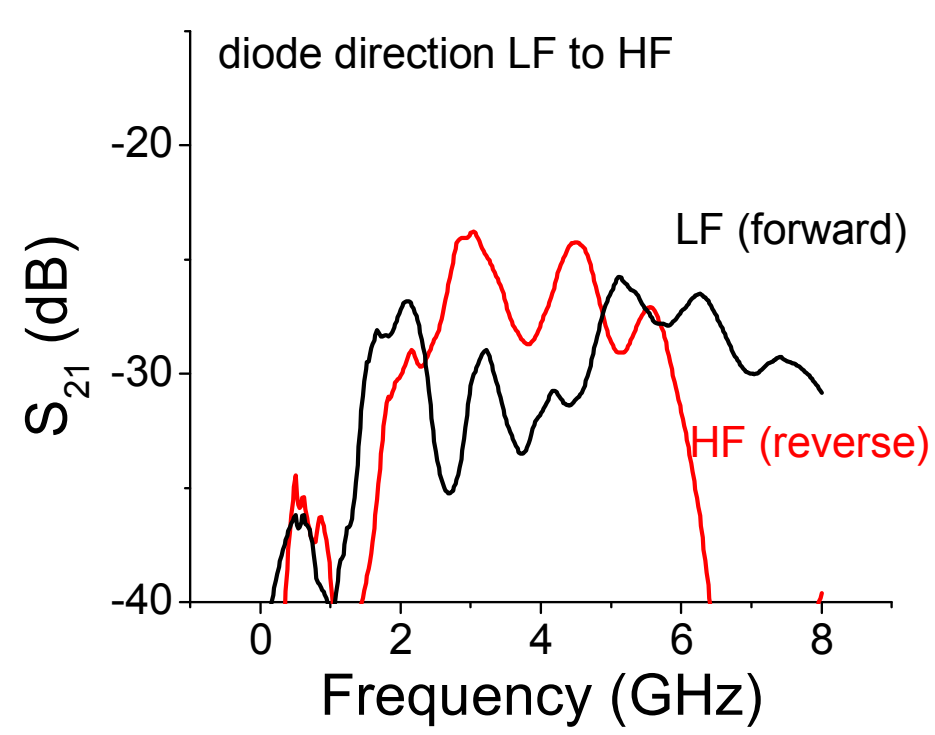

(a)

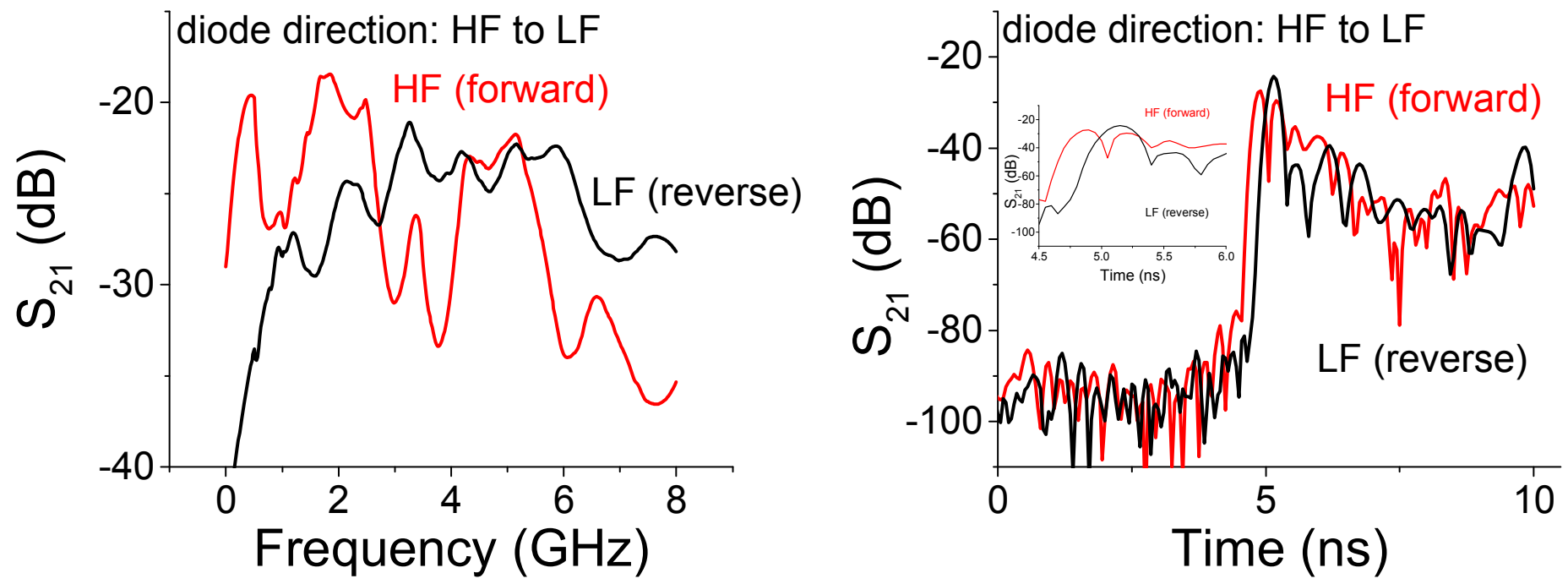

(b) (c)

**Fig. 12.** Experiments. (a) Transmission to either loop when the diode is forward biased from LF to HF. (b,c) The diode is forward biased from HF to LF; the overall frequency bandwidth has increased to 3 GHz in the 3-6 GHz region. The inset in (c) accentuates the temporal response from the HF loop (red) vs the very narrow output from the LF loop (black) of less than 150 ps.

## IV. Discussions and Conclusions

We analyzed two planar loops, one placed within each other, and coupled by a short, a diode, or the gap between them was left open. Their natural resonance frequency of one loop (HF) was twice the natural frequency of the other loop (LF). We divide the discussion into the simulation and experimental parts.

Simulations – frequency and time domains:
As explained, the model was run in the time domain using a monocycle that contained two lobes - negative and positive lobes. The complex amplitudes were than transformed to the frequency domain by Fast Fourier Transform (FFT). These simulations served as guide to the experimental data (using a VNA) which were carried in the frequency domain and later transformed to the time domain (via IFFT) to ascertain coherence.

*Input through one loop; received by the remaining port:* The simulation of the input signal to the HF loop (and received by the LF loop, or HF/LF) for the open and short cases has a bandwidth of ca 1 GHz. The peak was situated at the 0.75 GHz, the resonance of the LF loop. When the input was fed to the LF port, the signal was split into two shifted peaks, at 0.5 and 1.25 GHz. peaks. On the receiving end, the signal has split into three peaks in the 1 to 1.5 GHz region with the addition of a peak at 0.25 GHz. One may note that the peak at 0.5 GHz has been shifted to 1 GHz.

Having a diode coupler, exhibited a single input peak at 0.75 GHz and asymmetrical output: HF/LF exhibited two peaks at 0.25 and 1.25 GHz and LF/HF exhibited a monotonously decreasing curve. Unlike the open or short cases, the signal for the diode case in the time domain has stabilized rather fast and did not show oscillations. Addition of a floating metal frame did not change the results.

*Input across the diode; received by either ports:* Here, the signal is affected by the diode polarity. Adding a series resistor did not change the received signal. The HF loop exhibited three peaks: at 0.75, 1.3 and 2.1 GHz. The LF loop exhibited two peaks in that region: 0.75 and 1.8 GHz. However, when the polarity of the diode was reversed (LF to HF) a shift in the loop's spectra is detected: for the HF loop, the 0.75 and its second harmonics 1.3 GHz peaks get closer; for the LF loop, the 0.75 and the third harmonics at 1.8 GHz become more separated.

Experiments – frequency and time domain:
The experiments were conducted in the frequency domain and the results were converted to the time domain via built in IFFT. For the diode coupled HF/LF case the frequency bandwidth has extended to 1.5 GHz (Fig. 11.a) and the time response decreased to less than 150 ps. A further increase in the frequency bandwidth occurs when the signal is fed across the diode:

(*Case 1*) When the diode polarity is from LF to HF (Fig. 12a.), the frequency bandwidth is quite small.

(*Case 2*) When the diode polarity is from the HF loop toward the LF loop (Figs. 12b and 12c) an large bandwidth is recorded for the LF port (for which the diode is reversed) - this higher-harmonic 3-6 GHz band is associated with a quite narrow temporal response of less than 150 ps.

(*Case 3*) When the signal is received by the HF port we observe two bands, centered at 2 and 5 GHz, respectively, at width of ca 1.1 GHz, in addition to a narrower peak at 0.4 GHz. The associated temporal response is a bit larger than in Case 2, extending through two lobes in the time domain.

In summary, coupled planar loops were analyzed and measured for their extended frequency conversion. In contrast to a thin loop resonator that exhibits a narrow frequency bandwidth of ca 150 MHz, we have experimentally demonstrated a coherent frequency bandwidth of ca 3 GHz, resulting in less than 150 ps temporal response. The effect was attributed to higher-harmonics excitations. Elements such as these could be advantageous for ultra-wide bandwidth communication, radar and non-linear quantum information systems at the near-field.